\documentclass[runningheads]{llncs}
\usepackage{graphicx}
\usepackage{tikz}
\usepackage{tkz-graph}
\usepackage{bbm}
\usepackage{booktabs}
\usepackage{listings}
\usepackage{algpseudocode}
\usepackage{algorithm}
\usepackage{amsfonts}
\usepackage{amsmath}
\usepackage{wrapfig}
\usepackage{pgfplots}
\usepackage{pgfplotstable}
\usepackage{upgreek}
\usepackage{amssymb}
\usepackage{graphicx}
\usepackage{amsmath}
\usepackage{url}
\usepackage{enumitem}
\usepackage{eurosym}
\usepackage{mymacros}
\usepackage{pict2e,picture}
\usepackage{bm}
\usepackage{fancyvrb}
\usepackage{blindtext}

\usepackage{graphicx} 
\usepackage{booktabs} 
\usepackage{xparse}   
\NewDocumentCommand{\rot}{O{45} O{1em} m}{\makebox[#2][l]{\rotatebox{#1}{#3}}}%

\usepackage{subfig}

\usepackage{listings}
\usepackage{xcolor}

\usepackage{scalerel,amssymb}

\colorlet{punct}{red!60!black}
\definecolor{background}{HTML}{EEEEEE}
\definecolor{delim}{RGB}{20,105,176}
\colorlet{numb}{magenta!60!black}

\lstdefinelanguage{json}{
    basicstyle=\normalfont\ttfamily,
    numbers=left,
    numberstyle=\scriptsize,
    stepnumber=1,
    numbersep=8pt,
    showstringspaces=false,
    breaklines=true,
    frame=lines,
    backgroundcolor=\color{background},
    literate=
     *{0}{{{\color{numb}0}}}{1}
      {1}{{{\color{numb}1}}}{1}
      {2}{{{\color{numb}2}}}{1}
      {3}{{{\color{numb}3}}}{1}
      {4}{{{\color{numb}4}}}{1}
      {5}{{{\color{numb}5}}}{1}
      {6}{{{\color{numb}6}}}{1}
      {7}{{{\color{numb}7}}}{1}
      {8}{{{\color{numb}8}}}{1}
      {9}{{{\color{numb}9}}}{1}
      {:}{{{\color{punct}{:}}}}{1}
      {,}{{{\color{punct}{,}}}}{1}
      {\{}{{{\color{delim}{\{}}}}{1}
      {\}}{{{\color{delim}{\}}}}}{1}
      {[}{{{\color{delim}{[}}}}{1}
      {]}{{{\color{delim}{]}}}}{1},
}

\usepackage[activate={true,nocompatibility},final,tracking=true,kerning=true,spacing=true,factor=1100,stretch=10,shrink=10]{microtype}

\usepackage[font=footnotesize,labelfont=bf]{caption}

\begin{document}

\title{An Event Data Extraction Approach from SAP ERP for Process Mining}
\titlerunning{An ED Extraction Approach from SAP ERP for PM}

%
\author{Alessandro Berti\inst{1,2} \and
Gyunam Park\inst{1} \and 
Majid Rafiei\inst{1} \and
Wil van der Aalst\inst{1,2}}
%
%
\institute{Process and Data Science Group (PADS), RWTH Aachen University, Germany \and Fraunhofer Gesellschaft, Institute for Applied Information Technology (FIT),\\ Sankt Augustin, Germany}
\maketitle              
\begin{abstract}
The extraction, transformation, and loading of event logs from information systems is the first and the most expensive
step in process mining. 
In particular, extracting event logs from popular ERP systems such as SAP poses major challenges, given the size and the structure of the data. 
Open-source support for ETL is scarce, while commercial process mining vendors maintain connectors to ERP systems supporting ETL of a limited number of business processes in an ad-hoc manner.  
In this paper, we propose an approach to facilitate event data extraction from SAP ERP systems.
In the proposed approach, we store event data in the format of object-centric event logs that efficiently describe executions of business processes supported by ERP systems.
To evaluate the feasibility of the proposed approach, we have developed a tool implementing it and conducted case studies with a real-life SAP ERP system.
\end{abstract}

\section{Introduction}
\label{sec:introduction}

Process mining is a branch of data science including techniques to discover process models from event data, so-called \textit{process discovery}, check the compliance of data against the process models, so-called \textit{conformance checking}, and enhance process models with constraints/information coming from the event logs, so-called \textit{enhancement}.
Such techniques have been adopted by various domains, including healthcare, manufacturing, and logistics.
The first step of applying the techniques is to extract event logs from the target information systems, e.g., Enterprise Resource Planning (ERP) systems. 
This usually requires a connection to the database(s) supporting the information system. 
Afterward, the extracted event log undergoes pre-processing steps to resolve various data quality issues, including incomplete information, noise, etc.
These steps are usually called ETL (Extraction, Transformation, and Load). 
The ETL phase is usually the most time-consuming part of a process mining project \cite{DBLP:conf/caise/EckLLA15}.

ERP systems contain valuable data based on which process mining techniques provide insights regarding the underlying real-life business processes. 
In particular, the SAP ERP system has a significant share in the ERP market (22.5\% in 2017, Gartner).
Extracting data from an SAP ERP system is particularly challenging as it involves many different tables/objects.
Due to its complexity, support to extracting event data from the SAP ERP system has only been limited to commercial vendors, e.g., Celonis and ProcessGold, which requires extensive interaction with domain experts. 
Moreover, the logs extracted by such extractors suffer from convergence/divergence problems \cite{DBLP:conf/sefm/Aalst19}.
This is due to the necessity to specify a \emph{case notion}. A case notion is a criteria to group events that belongs to the same execution of a business process. In ERP systems, different case notions can be used for the same data. For example, in a procure-to-pay process, we could specify as case notion the order, the single item of the order, the delivery, the invoice, or the payment.

This paper proposes a novel approach to guide and ease the extraction of event logs from SAP ERP.
The approach consists of two phases, i.e., 1) building \textit{graph of relations} and 2) extracting object-centric event logs.
We propose to use Object-Centric Event Logs (OCEL) as intermediate storage to collect the events extracted from different tables. OCEL does not require the specification of a case notion. 
Therefore, it provides flexible and comprehensive event data extraction.
OCEL can be used with Object-Centric Process Mining (OCPM) techniques or flattened to traditional event logs by selecting a case notion out of objects.
The proposed approach has been implemented as a prototypical extractor and evaluated using an SAP ERP system. 

The rest of the paper is organized as follows. Section \ref{sec:background} presents some background knowledge.
Section \ref{sec:approach} presents the proposed approach. 
Section \ref{sec:tool} presents a prototypal software implementing the ideas proposed in this paper.
Section \ref{sec:assessment} evaluates the processes extracted by the prototypal software on top of an educational SAP instance.
Section \ref{sec:relatedWork} presents the related work on extracting and analyzing event logs from SAP.

\section{Background}
\label{sec:background}

This section presents some background knowledge on OCEL, convergence/divergence problems, and SAP systems.

\subsection{Object-Centric Event Logs}
\label{sec:backgroundObjCentrLog}

Traditional event logs in process mining have events associated with a single case/process execution. These event logs, extracted from information systems, suffer from convergence/divergence problems \cite{DBLP:conf/sefm/Aalst19}. 
We have a \emph{convergence} problem when the same event is duplicated among different instances. This happens, for example, in an order-to-cash process, when \emph{item} is considered as the case notion, and an event of order creation can be associated with several items.
We have a \emph{divergence} problem when several instances of the same activity happen in a case while not being causally related. This happens, for example, in an order-to-cash process, when \emph{order} is considered as the case notion, and several instances of the same item-related activity are contained in the same order. 

OCEL relax the assumption that an event is associated with a single case. Instead, in an OCEL an event can be related to several objects, where every object is associated with a type. This results in a more natural way to extract event data from a modern information system. For example, in ERP systems, the event of order creation can currently involve an order document and several items. This resolves the convergence problem (since we do not need to duplicate the events anymore) and the divergence problems (since activities related to items of an order are not associated with the case of the general order).

Recently, the OCEL standard\footnote{\url{http://www.ocel-standard.org/}} has been proposed as the mainstream format for storing object-centric event logs~\cite{DBLP:conf/adbis/GhahfarokhiPBA21}. 
The format is supported by different implementations and libraries in various programming languages, e.g., Java (ProM framework) and Python.
OCEL can be used to discover object-centric process models~\cite{DBLP:journals/fuin/AalstB20,DBLP:conf/simpda/BertiA19}, which describe the lifecycle of different object types and their interactions.
Moreover, conformance checking can be done on multiple object types~\cite{DBLP:journals/fuin/AalstB20}.

\subsection{SAP: Entities and Relationships}
\label{sec:backgroundSap}

\begin{figure*}[!htb]
\centering
\includegraphics[height=0.3\textheight]{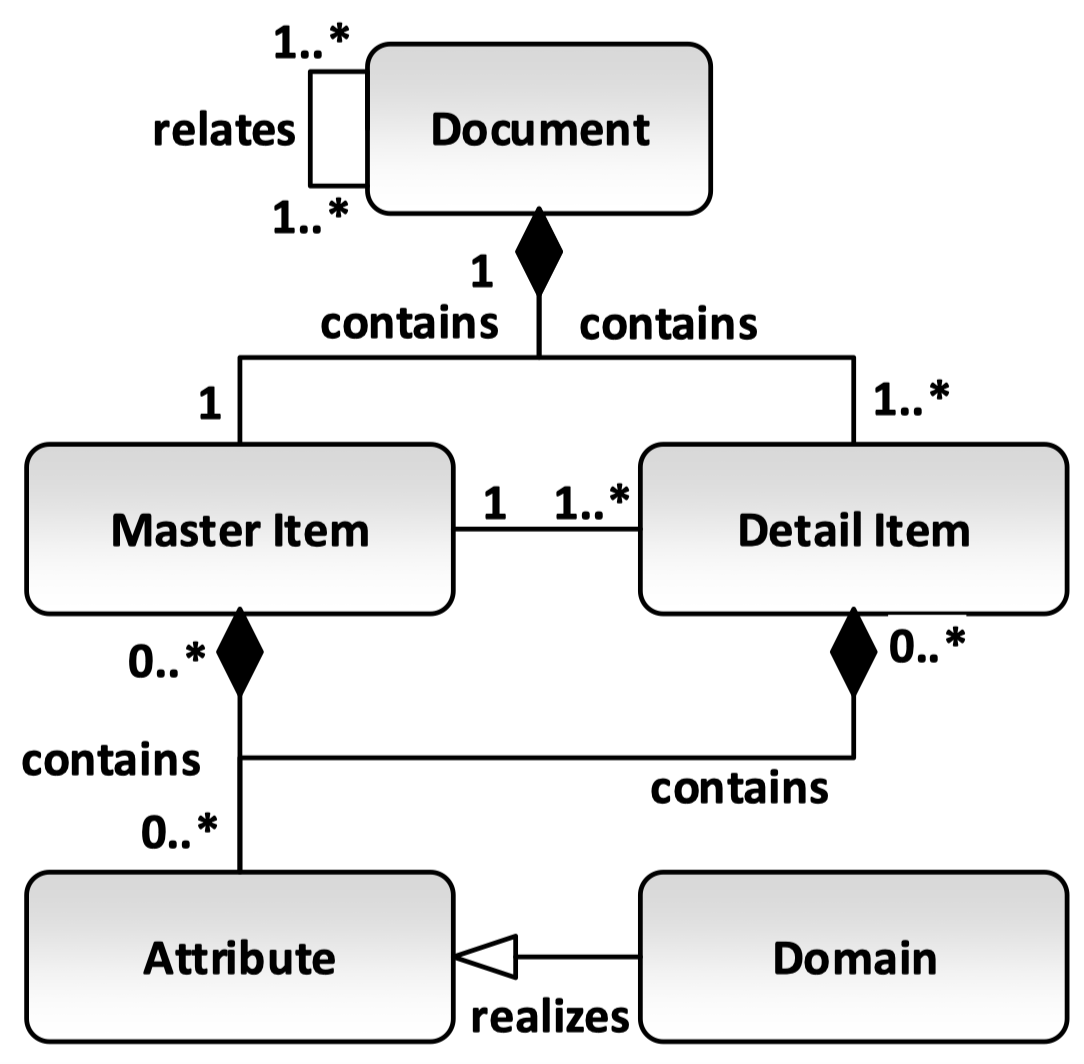}
\caption{Core entities of SAP ERP systems in UML 2.0 class diagram}
\label{fig:entireExtractionProcess}
\end{figure*}

In a broader sense, SAP ERP can be seen as a document management system. Therefore the concept of \emph{document}
is particularly important.
Fig. \ref{fig:entireExtractionProcess} introduces the document and its relevant entities and relationships among them, using UML 2.0 class diagram.
First, a document represents a core business object, including orders, deliveries, and payments.
Each document contains a  \emph{master item} and  \emph{detail items}.
For instance, a delivery document contains a delivery master item, corresponding to an order, and multiple delivery detail items, corresponding to materials in the order. 
A {\it master table} is a collection of the same type of master items, whereas a {\it detail table} is a collection of the same type of detail items.
For instance, {\it EKKO} as a master table contains purchase order master items.
{\it EKPO} as a detail table contains purchase orders detail items.

Both master and detail items contain a various number of \emph{attribute values}, e.g.,  the total cost of a document or the cost of a single item.
Each attribute belongs to a {\it domain} that encodes the type of information reported by the attribute, e.g., creation date and posting date of a document share the same domain because they are both dates.

\section{Extracting Event Data from SAP ERP: Approach}
\label{sec:approach}

\begin{figure*}[!b]
\centering
\includegraphics[width=\textwidth]{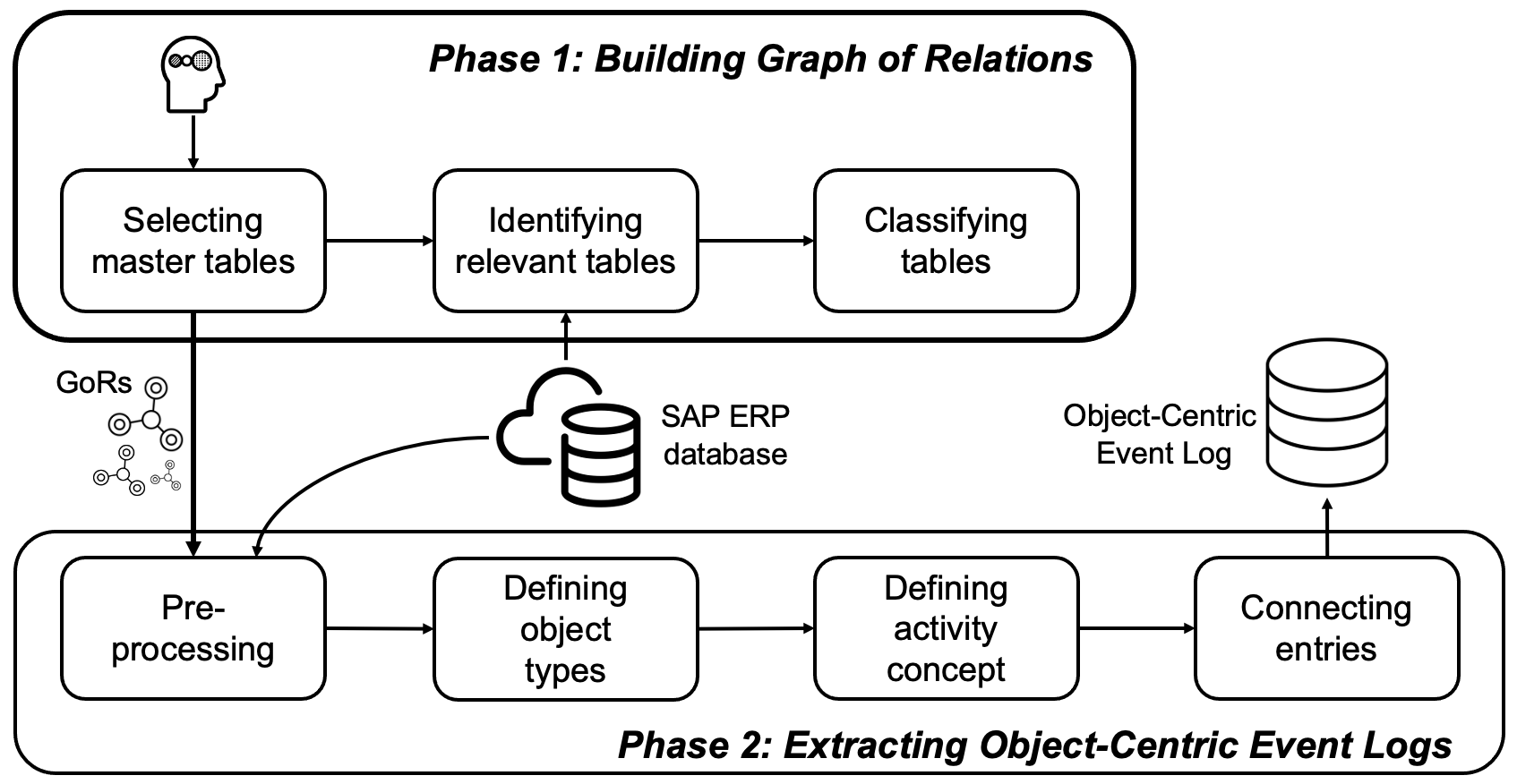}
\caption{Overview of extracting object-centric event logs from SAP ERP systems}
\label{fig:overview}
\end{figure*}

Figure \ref{fig:overview} describes an overview of our proposed approach to extract OCEL from SAP ERP systems. 
It consists of two phases: 1) building \textit{Graph of Relations (GoR)} and 2) extracting OCEL.
The former aims to construct a graph that describes all relevant tables of a business process.
There are well-known business processes in SAP ERP, e.g., Purchase to Pay (P2P) and Order to Cash (O2C).
For such business processes, target tables, where we extract event data regarding the process, are already known, e.g., \textit{EKKO}, \textit{RBKP}, \textit{EKBE} for P2P and \textit{VBAK}, \textit{BKPF} for O2C.
However, most business processes in an organization are mostly unknown and, thus, require the identification of relevant tables.

Based on the GoR, we extract OCEL by connecting them to the underlying database of SAP ERP systems.
To this end, we first preprocess records of tables described in the GoR. 
Next, we define activity concepts relevant to the target business process using the relevant tables.
Finally, based on the activity concept, we extract event data from the relevant tables.

\begin{figure*}[!b]
\centering
\includegraphics[width=\textwidth]{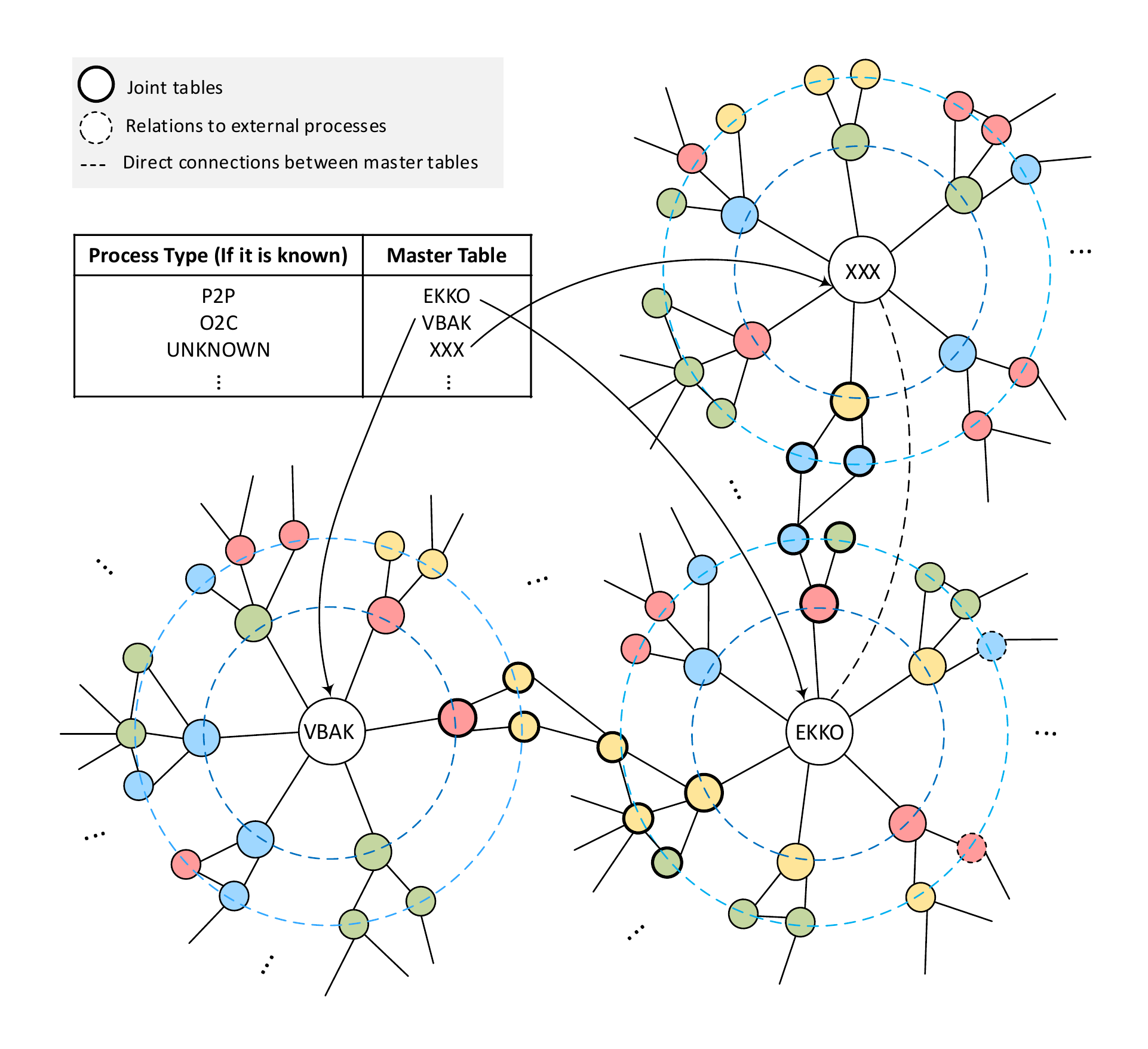}
\caption{Conceptual model of Graph of Relations (GoRs)}
\label{fig:concept}
\end{figure*}

\subsection{Building Graphs of Relations}\label{subsec:concept}

Figure~\ref{fig:concept} shows the conceptual model of three GoRs, each of which corresponds to a business process.
A GoR is an undirected connected graph where the nodes are SAP tables containing the potentially interesting information and the edges show a relation among two tables based on a joint field/column.  
The node in the center of a GoR is a master table that is most relevant to the target process.
The distance of each node from the master table shows the relevancy of the information contained in the corresponding table to the tables of interest and consequently to the corresponding type of process.
Different colors in a GoR indicate different classes of tables. 
Each class has a unique way of defining activity concepts. As a result, different GoRs may be connected to each other. Below are the steps to construct GoRs:

\subsubsection{Selecting Master Tables}
A GoR is built upon a master table relevant to a business process to analyze.
In this work, we consider relevant master tables as users' input. 

\subsubsection{Identifying Relevant Tables}
Based on the given master table, we need to identify relevant tables to the master tables. Such tables become the candidates for constructing the GoR. Three different main approaches may be taken: \textit{manual}, \textit{automatic}, and \textit{hybrid}.
\begin{itemize}
    \item In the manual approach, the identification is conducted by domain experts who understand business processes and the technical details of SAP systems. In addition, the domain expert may provide a data schema to explain the entities and relationships among them.
    \item In the automatic approach, the identification is made automatically by exploiting existing information in the system. For instance, using the table \textit{DD03VV}, one can extract the relationships between the tables.
    \item Finally, the hybrid approach exploits both manual and automated techniques. For instance, the data schema from domain experts can provide an initial set of relevant tables, which will be improved by including more relevant tables with the help of automatically generated relationships.
\end{itemize}

\subsubsection{Classifying Tables}
The last step is the classification of the identified tables into different classes.
In the following, we describe five different classes.
\begin{itemize}
    \item A \emph{flow table} describes the status of objects that compose the target business process. It explains the creation, deletion, and update of such objects, e.g., VBFA explains the status of objects that are associated with the Order-to-Cash (O2C) process. 
    \item A \emph{transaction table} describes the execution of transactions (TCODE) in SAP systems.
    \item A \emph{change table} describes the changes in objects of the target business process, e.g., CDHDR and CDPOS are primary change tables.
    \item A \emph{record table} stores relevant attributes of objects of the target business process, e.g., the table EKKO contains the relevant attributes of purchase order documents. 
    \item A \emph{detail table} stores the relationships between different entities, e.g., the table EKPO stores the connection between purchase requisitions and purchase orders.
\end{itemize}

\subsection{Extracting Object-Centric Event Logs}\label{subsec:extracting}

In this subsection, we explain how OCEL are extracted using GoRs. The extraction consists of four main steps; \textit{pre-processing}, \textit{defining activity concept},
\textit{defining object types}
and \textit{connecting entries}.

\subsubsection{Pre-processing}
SAP tables contain a lot of data related to different companies or groups in the same company (multi-tenant system). Moreover, when invoicing/accounting tables are considered, documents are organized by their fiscal year.
A pre-processing step must be performed to extract an event log of reasonable size, containing the desired behavior and a coherent set of information since document identifiers can be replicated across different organizations. To this end, the union of all the fields in the primary keys of the tables is considered, and for some of them, a filtering query is executed, e.g., on a specific company code or a specific fiscal year.

\subsubsection{Defining Object Types}
During the extraction, the entries of the master tables are transformed into events, having the columns as event attributes. Moreover, the values of all the columns except the dates and the numbers become objects of the object type given by the column's name.

\subsubsection{Defining Activity Concept}
To extract event data from GoRs, we take a divide-and-conquer approach. We first extract event data from each table and then combine them. The first step of extracting event data from each table is to define the activity concept. In the following, we explain how the activity is defined in each class of tables.
\begin{itemize}
    \item Each row in flow tables contains a current document number, a previous document number, the type of the current document, and the type of the previous document. For instance, considering VBELN as the domain, VBELN, VBELV, VBTYP\_N, and VBTYP\_V in the VBFA table contain respectively the current document number, the previous document number and the current and previous document types. We define activities as the type of the current documents, i.e., the value in VBTYP\_N.
    \item Each row in transaction tables contains a transaction code. We transform the transaction code into human-readable formats using the TSTCT table, e.g., VA02 is transformed to Change Order, which becomes the activity name.
    \item Each row in record tables describes the properties of an object. All the rows of the record tables are associated with the same activity, e.g., \emph{Create document [...]} for all the rows in EKKO.
    \item For change tables, we suggest three approaches: (1) Transaction codes used for changes are transformed into activities, (2) Fields, updated after changes, are converted into activities, e.g., \textit{Price Changed}, and (3) We consider both old and new fields' values and define activities, e.g., \textit{Postpone Delivery}, by comparing old and new values of delivery dates.
\end{itemize}

\subsubsection{Connecting Entries}
In this step, the information of the detail tables is used to enrich events. 
For example, if an entry of the table RSEG, containing detailed information about invoices, associates an invoice identifier with an order identifier, every event associated with the invoice identifier is also associated with the order identifier in the subsequent step.

\section{Extracting Event Data from SAP ERP: Tool}
\label{sec:tool}

We implemented a tool in the Python3 language, available in the Github repository; \url{https://github.com/Javert899/sap-extractor}. The tool
is available as a web application implemented using the Flask framework and can be launched with the command;
\emph{python main.py}. The web application can be accessed at the address; \url{http://localhost:5000/new_extractor.html}. First, the extractor asks
the parameters of connection to the database supporting the SAP ERP instance. Then, it provides both a list of object classes contained
in the database and a list of pre-configured sets of tables related to the mainstream processes. The next step is the construction of the GoR,
which permits extending the set of tables. The following step is about pre-providing the values for the primary keys of the included tables, e.g., the client used during the connection and the fiscal year.
After this step, the identification of the type of tables and the extraction occurs, which permits obtaining an OCEL,
that can be flattened to a traditional event log or analyzed using object-centric techniques such as the ones provided in \url{https://github.com/Javert899/sap-extractor}.

\begin{figure*}[!t]
\includegraphics[width=\textwidth]{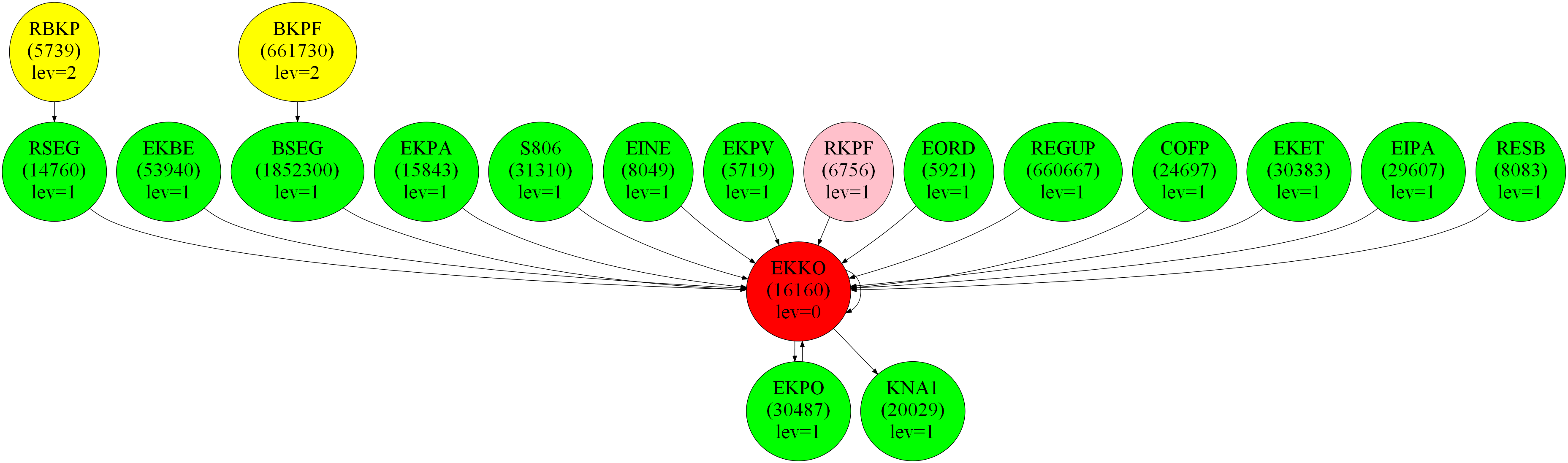}
\caption{A GoR built on our SAP IDES instance on the P2P process. Detail tables are colored by green, RKPF that is an additional record tables is colored by pink, and RBKP and BKPF that are additional transaction tables are colored by yellow.}
\label{fig:gorP2P}
\end{figure*}

\section{Assessment}
\label{sec:assessment}

This section proposes an assessment of the proposed techniques on top of an SAP ERP IDES system. In particular, we will target the extraction of the well-known Purchase to Pay (P2P) system.
A P2P process involves different steps including \textit{approval of a purchase requisition}, \textit{placement of a purchase order}, \textit{invoicing from a supplier}, and \textit{payment}.
Therefore, it involves different tables in the SAP system.

\subsection{Building a Graph of Relations}

\subsubsection{Selecting Master Tables}
The first step in the tool is selecting a candidate table related to the process. In this case, we start from EKKO that is one of the main tables in the P2P process and contains the master information. 
In building the GoR, represented in Fig. \ref{fig:gorP2P}, several other tables that are connected to EKKO are found.
Given the vast number of tables contained in SAP, we applied a simple filtering based on the number of entries in each table to show the main nodes in the GoR.

\subsubsection{Identifying Relevant Tables}
Figure \ref{fig:gorP2P} shows other tables containing event data meaningful to extract an event log for the P2P\footnote{Including EKBE, containing goods/invoice receipts, BSEG, containing detail table for payments, RSEG, containing detail data for invoices, RKPF, including inventory management data, EKPO, containing the detailed information about the purchase orders, EKPA, containing the partner roles in purchasing, and EKET, containing the scheduling agreement. We can see that the EBAN table, containing purchase requisition data, has not been included because of the filtering applied on the number of entries. However, it would be found by the method if
the threshold is set to a lower value so that we will include it in the following steps.}. The user needs to specify the tables to include along with the original set of tables.
The GoR is therefore updated\footnote{The set of tables to extract include: EKKO, EKPO, EKPA, EKET, EKBE, BSEG, BKPF, RSEG, RBKP, RKPF, RESB, EBAN.}. In our implementation, the master tables related to the detail tables are automatically included in the set\footnote{This means that BKPF, the master table of BSEG, containing the master data about the payments,
and RBKP, the master table of RSEG, containing the master data about the invoices, are included.}.

\subsubsection{Classifying Tables}
The tool needs to categorize the tables in the set between master tables and detail tables, as the master tables contain event data, while detail tables contain the connection between
different entities:
\begin{itemize}
\item Some tables are recognized as transactions tables: \emph{RBKP} (containing the transactions related to the invoices) and \emph{BKPF} (containing the transaction related to the payments).
\item Some tables are recognized as record tables: \emph{EBAN} (in which a record is a purchase requisition), \emph{EKKO} (in which a record is an order document), and \emph{RKPF} (in which a record is a reservation).
\item Some tables are recognised as detail tables: \emph{EKPO}, \emph{EKPA}, \emph{EKET}, \emph{EKBE}, \emph{BSEG}, \emph{RSEG}, \emph{RESB}\footnote{Because their primary key is contained in the primary key of \emph{EKKO} (for EKPO, EKPA, EKET, EKBE), \emph{BKPF}, \emph{RBKP} and \emph{RKPF} respectively.}.
\end{itemize}

\subsection{Extracting Object-Centric Event Logs}


In this section, we will explain the main steps of the log extraction process, including the definition of the
object types and the
activity concept for the extraction, and the connection between the entries given the information of the detail tables. Since we did not perform a pre-processing step, we will not assess the step here.

\subsubsection{Defining Object Types}

Starting from the choices on the GoR and the identification of the type of tables, it is possible to extract different object types,
including \emph{BANFN-BANFN} (purchase requisition), \emph{INFNR-INFNR} (purchasing record), \emph{EBELN-EBELN} (purchase order), \emph{BELNR-RE\_BELRN} (the invoice number), \emph{BELNR-BELNR\_D} (the payment number),
and \emph{AWKEY-AWKEY} (a generic object type containing the ID of the object in SAP).

\subsubsection{Defining Activity Concept}

The activity concept is defined as follows:
\begin{itemize}
\item For the record tables, a unique activity is defined for all the events, that is \emph{Create document (TABNAME)} (where TABNAME is the name of the corresponding record table, so it can be EBAN/EKKO/RKPF).
\item For the transaction tables, the activity is given by the transaction code\footnote{Using the description of the transaction contained in the table TSTCT.}. Mainstream transactions occurring are \emph{Enter incoming invoice}, \emph{Enter incoming payment}, \emph{Enter outgoing payment}.
\end{itemize}

\subsubsection{Connecting Entries}
The detail tables are used to enrich the entries extracted from the master tables as follows:
\begin{itemize}
    \item \emph{BSEG} provides a connection from the payments to the purchase order items.
    \item \emph{RSEG} connects the invoices to the purchase order items.
    \item \emph{EKPO} provides a connection of the purchase order items to the corresponding purchase requisition.
    \item \emph{EKPA} and \emph{EKET} contain detailed information that does not provide meaningful links to other tables in the set. \emph{EKBE} is a peculiar type of detail table, as it contains the information about goods/invoice receipts, so it could be seen as a master table. Still, it also links the purchase order items with the invoices through the goods/invoice receipts.
\end{itemize}


\section{Related Work}
\label{sec:relatedWork}

This section presents the related work on data extraction from ERP systems for process mining purposes.

\subsubsection{Data Extraction and Pre-Processing from SAP ERP}

In \cite{van2004process}, an approach to extract traditional event logs from SAP ERP is proposed.
The set of relevant business objects is identified, and the related tables and their relations are identified.
A limitation is that the construction
of the document flow is manual.
In \cite{DBLP:conf/bpm/IngvaldsenG07}, the authors address the pre-processing challenges to extract event logs from SAP ERP by using tools such as EVS Model Builder.
In \cite{DBLP:conf/bpm/CalvaneseMSA15}, an ontology-driven approach for the extraction of event logs from relational databases is proposed, in which the user can express semantic queries which are then translated to relational queries.
In \cite{DBLP:conf/bpm/JansS17}, the effects of some decisions on the quality of the resulting event log are analyzed. In particular, the context of event log extraction from ERP system is considered.

\subsubsection{Artifact-centric Models on ERP systems}

In \cite{DBLP:journals/tsc/LuNWF15}, an approach to discover artifact-centric models from ERP systems is proposed.
The approach is split into two main parts: 1) identifying a set of artifacts, extracting a traditional event log, and a model of its lifecycle;
2) discovering the interactions between artifacts.
The set of tables to extract needs to be decided by the user and the specification of the activity concepts is not described in this work.

In \cite{DBLP:conf/caise/LiMCA18}, object-centric event logs (in the XOC format) are extracted from the Dollibar ERP system. These logs have been used to generate an object-centric behavioral
constraints (OCBC) model. However, OCBC/XOC are not scalable.

\subsubsection{OpenSLEX Meta-Models}

In \cite{DBLP:conf/bpm/MurillasAR15}, a meta-model is proposed to ease the extraction of process mining event logs from information systems
supported by relational databases.
The instances of the OpenSLEX meta-model can be built from different types of database logs (redo logs, SAP change tables). Hence, the meta-model is generic and not tailored to the
peculiar features of an SAP ERP system. The main problem is that the extraction of an event log requires a case notion's specification, which leads to convergence/divergence problems.

\subsubsection{Enterprise-Grade Connectors}
Several commercial vendors of process mining solutions offer enterprise-grade connectors to SAP, that are able to ingest and process millions of events. Notable examples in the current landscape are Celonis\footnote{\url{https://www.celonis.com/solutions/systems/sap}}, Signavio\footnote{\url{https://www.signavio.com/products/process-intelligence/}}, LANA\footnote{\url{https://lanalabs.com/en/migration-to-sap-s-4-hana-with-lana/}}, UIPath\footnote{\url{https://docs.uipath.com/process-mining/docs/introduction-to-sap-connector}}.

\section{Conclusion}
\label{sec:conclusion}

In this paper, we proposed a generic approach to extract event logs from SAP ERP, which exploits the relationships between tables in SAP to build Graphs of Relations (GoRs) and obtains Object-Centric Event Logs (OCEL) using GoRs. Figure \ref{fig:overview} summarizes our approach.
By storing extracted event data into OCEL, we permit the specification of multiple case notions, avoiding the convergence/divergence problems and simplifying the extraction process.
An open-source tool implementing the approach and a case study on an educational SAP instance have been presented, showing the feasibility of identifying the relationships between different tables of the P2P process and extracting corresponding OCEL.
As future work, we plan to deploy our approach on different instances of SAP systems running in real businesses to explore the connection between GoRs and underlying processes and to discover unknown processes. Moreover, we should further assess how good the extraction of a typical SAP process is in comparison to commercial-grade extractors.

\bibliographystyle{splncs04}
\bibliography{extractor}

\end{document}